\documentclass[reprint,superscriptaddress,amsmath,amssymb,aps]{revtex4-1}

\usepackage{graphicx}% Include figure files
\usepackage{dcolumn}
\usepackage{bm}
\usepackage{mathrsfs}
\usepackage{mathtools}
\usepackage{balance}
\usepackage{hyperref}
\usepackage{filecontents}
\usepackage[USenglish]{babel}

\begin{document}

%%%%%%%%%%%%%%%%%%%%%%%%%%%%%%%%%%%%%%%%%%%%
\title{Influence of surface restructuring on the activity of SrTiO$_{3}$ photoelectrodes \\for photocatalytic hydrogen reduction}
%%%%%%%%%%%%%%%%%%%%%%%%%%%%%%%%%%%%%%%%%%%%

\author{Yihuang Xiong}
\email{Email:yyx5048@psu.edu}
\affiliation{Department of Materials Science and Engineering, Materials Research Institute, The Pennsylvania State University, University Park, PA 16802, USA}
\author{Ismaila Dabo}
\affiliation{Department of Materials Science and Engineering, Materials Research Institute, The Pennsylvania State University, University Park, PA 16802, USA}

\begin{abstract}
Perovskite photoelectrodes are being extensively studied in search for photocatalytic materials that can produce hydrogen through water splitting. The solar-to-hydrogen efficiency of these materials is critically dependent on the electrochemical state of their surface. Here, we develop an embedded quantum-mechanical approach using the self-consistent continuum solvation (SCCS) model to predict the relation between band alignment, electrochemical stability, and photocatalytic activity taking into account the long-range polarization of the semiconductor electrode under electrical bias. Using this comprehensive model, we calculate the charge-voltage response of various reconstructions of a solvated SrTiO$_{3}$ surface, revealing that interfacial charge trapping exerts primary control on the electrical response and surface stability of the photoelectrode. Our results provide a detailed molecular-level interpretation of the enhanced photocatalytic activity of SrTiO$_{3}$ upon voltage-induced restructuring of the semiconductor-solution interface.
\end{abstract}

\maketitle

\section{\label{sec:Introduction}Introduction}

Hydrogen is a sustainable energy carrier whose electrocatalytic reaction with oxygen produces electricity and heat without emitting carbon dioxide. A highly attractive approach for the production of hydrogen fuels consists of splitting water molecules by photocatalytic means; however, engineering photoactive electrode materials that can efficiently promote this reaction remains an outstanding question at both the experimental and theoretical levels. Strontium titanate (SrTiO$_{3}$) is a photocatalytic material that has shown promising solar-to-hydrogen performance \cite{doi:10.1021/jp953720e,BARD197959,doi:10.1021/ja00426a017,Mavroides1976}; under ultraviolet light, this wide-bandgap semiconductor exhibits a high quantum efficiency in converting incident photons into charge carriers \cite{doi:10.1021/ja00426a017,Mavroides1976}.

To date, considerable efforts have been dedicated to understanding the microscopic mechanisms that underlie the water-splitting performance of SrTiO$_3$ \cite{doi:10.1021/ja00426a017,Mavroides1976,Fujishima1972,Wagner1980}. A central aspect of these studies has been to elucidate the structure of the SrTiO$_{3}$-water interface. Through surface-sensitive characterization and electronic-structure calculation, it has been shown that SrTiO$_{3}$ can undergo a TiO$_{2}$-rich surface reconstruction \cite{Erdman2002,TiO2-rich,PhysRevLett.98.076102,Erdman2003,Kubo2003,Castell2002}. This result has been further confirmed by the detailed comparison of computationally predicted structures with accurate x-ray reflectivity data \cite{Schlom2016,PhysRevLett.98.076102}.

Beyond their descriptive power, density-functional theory simulations are now frequently applied to address many of the questions that surround the performance of water-splitting catalysts. These calculations have been used to evaluate the band edge positions against redox potentials in electrolytic media \cite{doi:10.1021/jacs.5b00798,PhysRevB.83.235301,Hormann2019}, elucidate catalytic reaction pathways \cite{doi:10.1021/jacs.6b04004,doi:10.1021/cs500668k,doi:10.1021/ja511572q,doi:10.1021/acscatal.6b01138}, and narrow down the choice of candidate photocatalysts \cite{C2EE23482C,castelli2015new}. Furthermore, it is now possible to achieve a microscopic understanding of the electrical conditions that exist in the subsurface depletion region of a photoelectrode through first-principles Mott-Schottky analysis, which enables one to capture the driving forces that drag or push the photogenerated charge carriers to the interface \cite{Andreussi2012,Campbell2017}.

Using these newly available computational models, we undertake here a detailed analysis of various reconstructions of the SrTiO$_{3}$ electrodes to predict and understand their photoelectrochemical properties. By simulating the effects of band bending and band alignment, we examine the influence of surface termination on the electrical response and electrochemical stability of SrTiO$_{3}$ under voltage to shed light into its photocatalytic performance as a function of preparation and operation conditions.

First, we outline the computational methods in Sec.~\ref{sec:computational_methods} with a presentation of the surface models and description of the voltage-dependent surface calculations. We then report computational predictions of voltage-induced charge accumulation and  band bending as a function of the surface termination in Sec.~\ref{sec:results}. Finally, we discuss the consequences of these predictions in understanding the electrochemical stability and photocatalytic activity of the reconstructed SrTiO$_{3}$ electrodes.

%%%%%%%%%%%%%%%%%%%%%%%%%%%%%%%%%%%%%%%%%%%%
\section{ Computational method}
\label{sec:computational_methods}
%%%%%%%%%%%%%%%%%%%%%%%%%%%%%%%%%%%%%%%%%%%%

%%%%%%%%%%%%%%%%%%%%%%%%%%%%%%%%%%%%%%%%%%%%
\subsection{First-principles simulations}
\label{sec: first-principles method and bulk structure}
%%%%%%%%%%%%%%%%%%%%%%%%%%%%%%%%%%%%%%%%%%%%
Self-consistent-field calculations are performed at the semilocal Perdew-Burke-Ernzerhof level \cite{Perdew1996} with the on-site Hubbard $U$ parameterization of the self-interaction correction to the effective potential using the {\sc pw} code of the {\sc quantum-espresso} distribution \cite{Giannozzi2009}. Ionic cores are represented by norm-conserving pseudopotentials with a kinetic-energy cutoff of 100 Ry for the reciprocal-space expansion of the wave functions. For the bulk structure of SrTiO$_{3}$, the Brillouin zone is sampled with a 6 $\times$ 6 $\times$ 6 Monkhorst-Pack grid \cite{Monkhorst1976}. The Hubbard $U$ correction has been shown to yield an improved description of electronic structures \cite{PhysRevB.71.035105}, magnetic orderings \cite{PhysRevB.99.094102,PhysRevB.85.085306} and catalytic properties \cite{doi:10.1021/jp306303y,doi:10.1021/ja405997s}. We thus employ this approach for simulating SrTiO$_3$ using a Hubbard parameter of  $U$ = 3.0 eV for the Ti sites, which has been computed self-consistently from linear-response theory \cite{PhysRevB.71.035105,PhysRevB.98.085127}. Through variable-cell relaxation, the lattice constant is predicted to be 3.90 {\AA} in good agreement with experiment (3.91 {\AA}). The bandgap of the optimized structures is calculated to be 2.4 eV, which is lower than the experimental bandgap of 3.2 eV \cite{PhysRev.140.A651} but significantly improved compared to the bandgap of 1.8 eV obtained without self-interaction correction. We note that an overestimated bandgap of 4.78 eV is predicted when the on-site Hubbard correction is applied to the oxygen 2$\textit{p}$ orbitals. Therefore, the Hubbard correction is only applied to Ti in all the calculations presented in Sec.~\ref{sec:results}.

%%%%%%%%%%%%%%%%%%%%%%%%%%%%%%%%%%%%%%%%%%%%
\subsection{Surface structure of slab models}
\label{sec: description of slabs surface }
%%%%%%%%%%%%%%%%%%%%%%%%%%%%%%%%%%%%%%%%%%%%

The TiO$_2$ surface termination has been reported to form when SrTiO$_{3}$ is annealed at 850-1,000 $^\circ$C \cite{Erdman2002,Erdman2003,Kubo2003,Castell2002}. Depending on the preparation methods, SrTiO$_{3}$ exhibits various surface structures along the (001) direction. Previously reported terminations include the single-TiO$_{2}$ layer \cite{CORD198534,PhysRevLett.98.076102,Castell2002}, and the stoichiometric double-TiO$_2$ terminated interface in the (1$\times$1) and (2$\times$1) surface unit cells (2 ML (1$\times$1) and 2 ML (2$\times$1)) \cite{Erdman2002,Erdman2003,PhysRevLett.98.076102,Castell2002}. In addition, a unique type of SrTiO$_3$ surface structure induced under electrochemical conditions has been recently reported by Plaza and coworkers \cite{Schlom2016}. Specifically, the SrTiO$_3$ interface has been shown to undergo a substantial reconstruction upon ``training'' at positive bias, forming a non-stoichiometric, triple-TiO$_2$ terminated interface, which exhibits significantly improved activities in alkaline solutions. We note that the SrO-terminated interface could also form under certain synthesis conditions \cite{COX1983247,Hussain}, and it has been shown theoretically to have thermodynamical stability comparable to that of the TiO$_2$ termination in vacuum \cite{Vanderbilt,Vanderbilt2}. It should be mentioned that there exist varying opinions regarding the stability of the SrO termination in aqueous environments \cite{COX1983247,Koichiro2003,Guhl,Nesbitt1981,Seitz1011}; although a careful study of the occurrence and (photo)electrochemical response of the SrO interface is of primary interest, its discussion is beyond the scope of the present work.

\begin{figure}[t]
 	\centering
 	\includegraphics[width=0.45\textwidth]{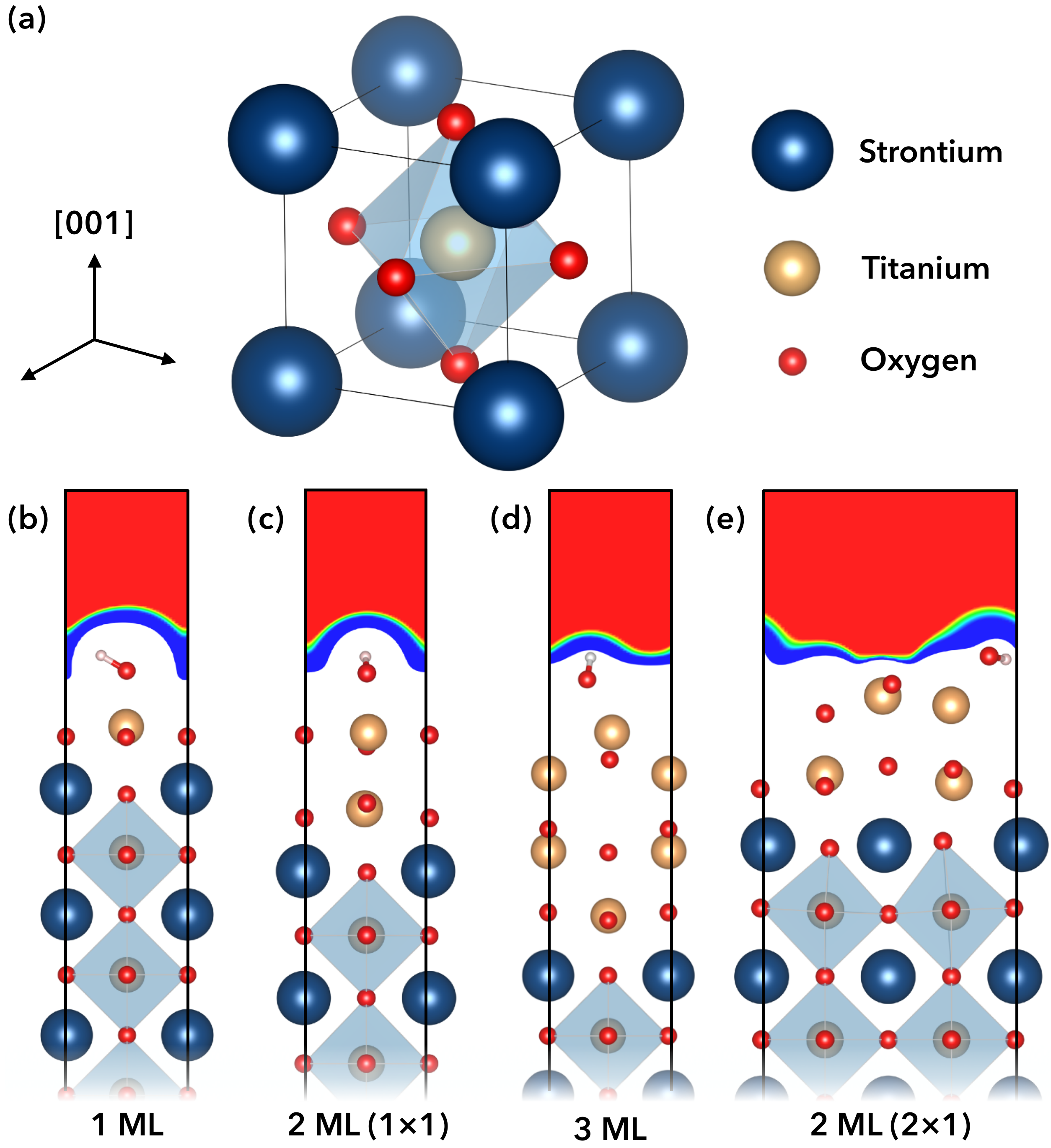}    
 	\caption{(a) Perovskite cubic structure of SrTiO$_{3}$. Schematics of interface structures that include (b) Single-TiO$_{2}$ termination (1 ML), (c) double-TiO$_{2}$-terminated (1$\times$1) reconstruction (2 ML (1$\times$1)), (d) triple-TiO$_{2}$ termination (3 ML) and (e) double-TiO$_{2}$-terminated (2$\times$1) reconstruction (2 ML (2$\times$1)) with hydroxyl terminations. All the surface terminations that are studied in this work are presented in detail in Fig.~\ref{fig2:Terminations}. The colored regions represent the continuum electrolyte.}
 	\label{Fig1.Surface_structure}
\end{figure}

\begin{figure*}
 	\centering
 	\includegraphics[width=0.8\textwidth]{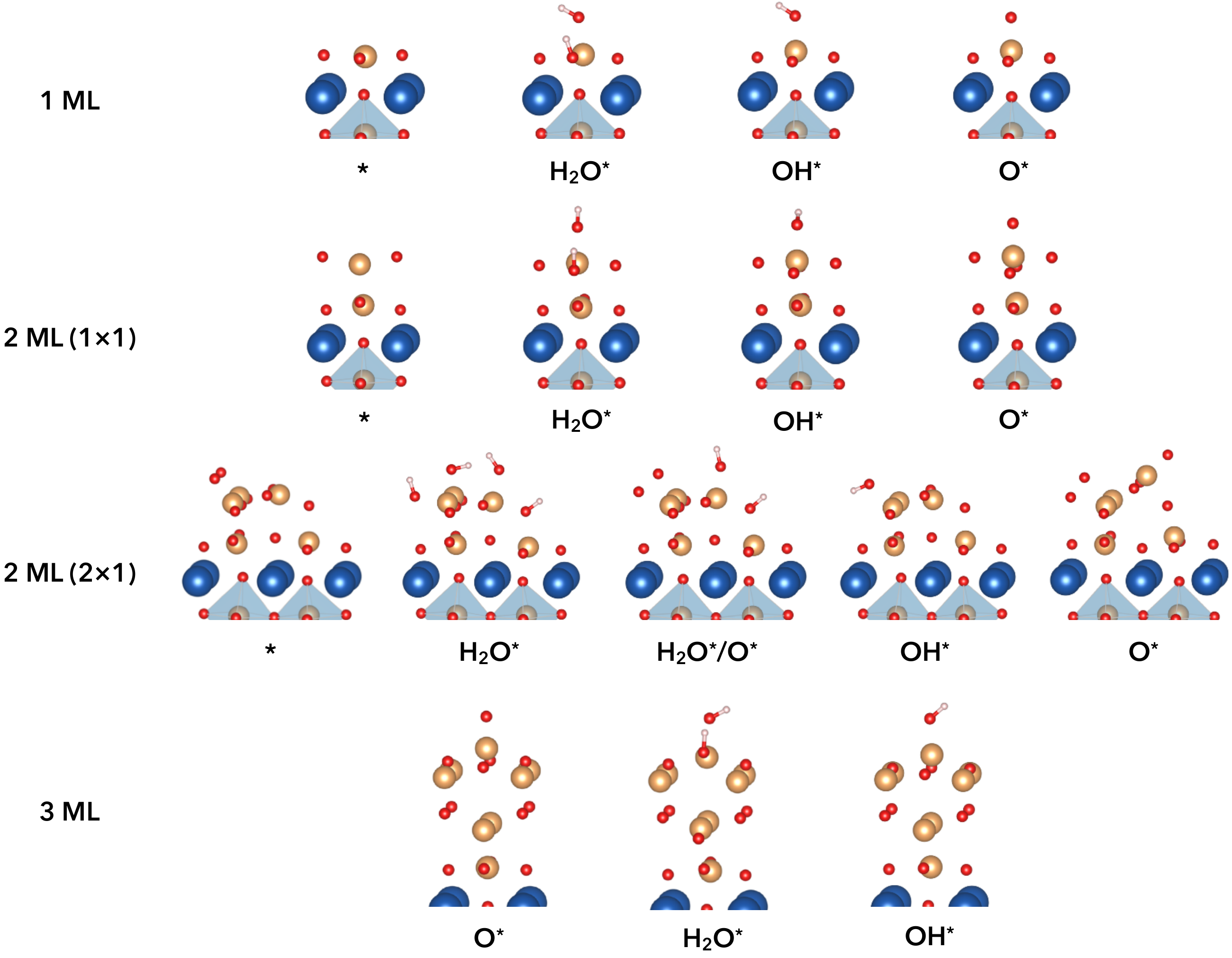}    
 	\caption{Various terminations of the SrTiO$_{3}$ interfaces are constructed by firstly hydrating the interfaces with dissociated water molecule(s), then progressively removing the proton(s) until the interfaces are fully oxygenated. The termination groups are labeled as *, H$_{2}$O$^{*}$, OH$^*$ and O$^*$ for the bare interface, water molecule,  hydroxyl group, and oxygen termination, respectively. For the 2 ML (2$\times$1) termination, H$_{2}$O$^*$/O$^*$ stands for the partial oxidation of the SrTiO$_3$-water interface.}
 	\label{fig2:Terminations}
\end{figure*}

Based on this structural survey, four symmetric slab models are constructed, including 7 bulk layers with a periodic separation of approximately 15 $\rm \AA$ along the transverse axis.
A schematic of the slab models is shown in Fig.~\ref{Fig1.Surface_structure}. Various terminations are built following Ref.~\cite{doi:10.1021/ja511332y}, where all the interfaces are firstly decorated with dissociated water molecules, by placing a hydroxyl group and a proton at the interfacial Ti and O sites, respectively. The protons are then progressively removed to reach fully oxygenated interfaces. All the terminations that are considered in this work are presented in Fig.~\ref{fig2:Terminations}. The Brillouin zone of each slab is sampled using a 6 $\times$ 6 $\times$ 1 grid of wave vectors and 0.01 Ry of Marzari-Vanderbilt cold smearing \cite{PhysRevLett.82.3296}. To retain the bulk characteristics of SrTiO$_{3}$, the middle 3 layers are fixed and other atoms are fully relaxed until interatomic forces are brought down to 0.01 eV $\rm \AA^{-1}$.

%%%%%%%%%%%%%%%%%%%%%%%%%%%%%%%%%%%%%%%%%%%%
\subsection{Solvation effects}
\label{sec:Solvation_method}
%%%%%%%%%%%%%%%%%%%%%%%%%%%%%%%%%%%%%%%%%%%%

To describe the solvation environment, the structure is immersed in an implicit polarizable solvent parameterized by the self-consistent continuum solvation (SCCS) model \cite{Andreussi2012}. In this model, the shape of the dielectric cavity is defined self-consistently from the electron density of the solvated surface that is directly computed at the quantum-mechanical level. The SCCS model has been shown to efficiently capture the essential features of liquid-solid interface through its logarithmically definition of the solvation shell: $\epsilon(\rho) = {\rm exp}[(\zeta - {\rm sin}(2\pi \zeta)/2\pi){\rm ln}\epsilon_{\rm s}]$, which involves the smooth switching function $\zeta(r) = (\rm{ln}\rho_{\rm max}-\rm{ln}\rho)/(\rm{ln}\rho_{\rm max}-\rm{ln}\rho_{\rm min})$ that defines the gradual dielectric transition. $\rho_{\rm max}$ and $\rho_{\rm min}$ denote the thresholds of the electron density that define the frontiers of solute ($\epsilon = 1$) and solvent ($\epsilon = \epsilon_{\rm s}$), respectively. Non-electrostatic cavitation contributions including surface tension, external pressure, dispersion and repulsion interactions are also incorporated into the SCCS model. These contributions are explicitly expressed as $F_{\rm cav}= \gamma S$ and $F_{\rm disp+rep} = \alpha S + \beta V$, where $\gamma$ stands for the surface tension of the solvent, $\alpha$ and $\beta$ are parameterized against experimental solvation energies, and $V$ and $S$ are the quantum volume and quantum surface area that are defined as $V = \int \Theta d{\boldsymbol{r}}$ and $S = -\int d \Theta / d\rho |{\nabla \rho}|d{\boldsymbol{r}}$ using the additional switching function $\Theta(\rho) = (\epsilon_{\rm s}-\epsilon(\rho))/(\epsilon_{\rm s} -1)$. Specifically, the parameters are: $\epsilon_{\rm s}$ = 78.3 is the dielectric constant of water, $\rho_{\rm min}$ = $10^{-4}$ a.u., $\rho_{\rm max}$ = $5 \times 10^{-3}$ a.u., $\gamma$ = 72.0 dyn/cm, $\alpha$ = -- 22 dyn/cm, and $\beta$ = --0.35 GPa. It has been discussed recently that the introduction of the volume-dependent energy term is unphysical for a surface system \cite{doi:10.1021/acs.jctc.7b00375,Andreussi2019}. The cavitation energy on the other hand provides a minor improvement in the accuracy of the results \cite{Andreussi2019,Andreussi2012}, and would largely cancel out for slab setups \cite{Huang2018}. Therefore, those terms are not included in this work.

%%%%%%%%%%%%%%%%%%%%%%%%%%%%%%%%%%%%%%%%%%%%
\section{Results and discussion}
\label{sec:results}
%%%%%%%%%%%%%%%%%%%%%%%%%%%%%%%%%%%%%%%%%%%%

\begin{figure}[t]
    \hspace*{-0.8cm}   
	\includegraphics[width=0.57\textwidth]{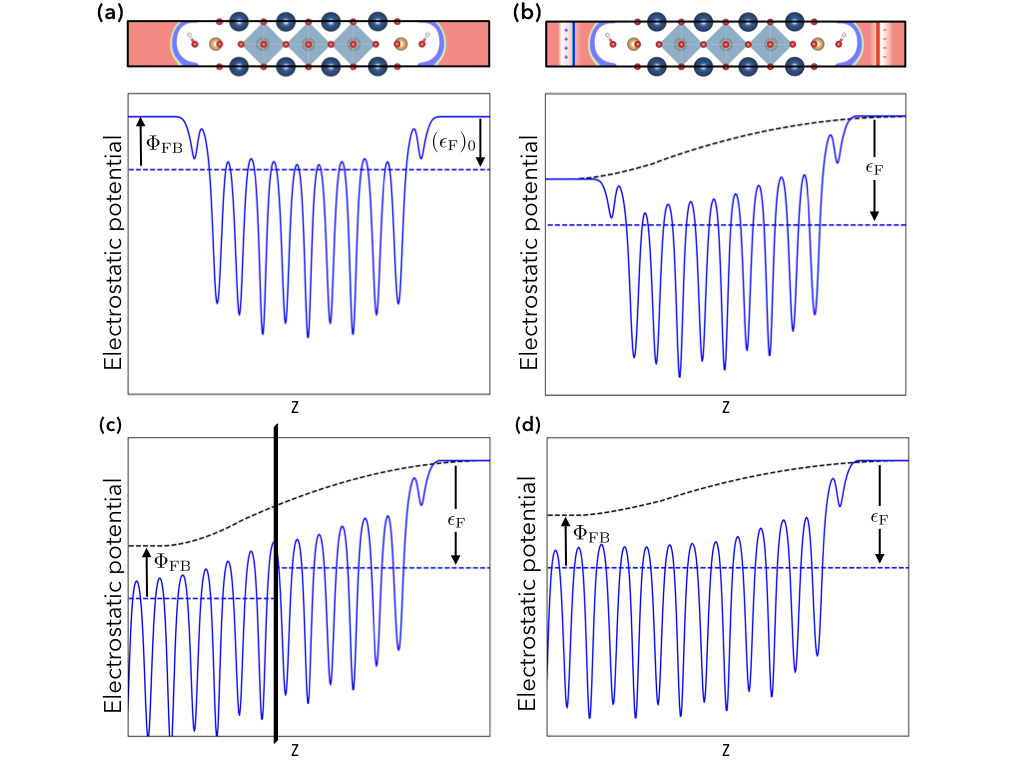}    
	\caption{\small Band bending at the SrTiO$_{3}$-electrolyte interface is accounted for by biasing the original electrostatic potential along transverse $z$ coordinate of the neutral slab (panel a) with explicit charge in the electrode ($\Phi_{\rm FB}$ denotes the opposite of the Fermi energy $(\epsilon_{\rm F})_0$ (per electron) with respect to asymptotic electrostatic reference). This is achieved by placing Helmholtz layers of countercharges at the interfaces while maintaining charge neutrality (panel b). A cutoff plane located at the inflection point of the difference between the charged- and neutral-slab potentials (shown by the upper dashed curve) defines the onset of the Mott-Schottky extrapolation of the potential inside the electrode (panel c). Finally, the Fermi levels of the bulk and interface are matched by varying the explicit charge on the electrode (panel d).}
	\label{fig3:alignment}
\end{figure}

To derive the charge-voltage characteristics of the SrTiO$_{3}$-water interfaces, we employ the procedure initially proposed by Campbell and coworkers \cite{Campbell2017} to convert a finite slab into a semi-infinite surface capturing the bending of the electronic bands in the subsurface layers. This procedure is depicted schematically in Fig.~\ref{fig3:alignment}; two planar countercharges are placed 3 $\rm \AA$ away from both ends of the SrTiO$_{3}$ surface to represent the Helmholtz contribution to the polarization of the interface, then the electrode is partitioned into an explicit finite interface region and an implicit semi-infinite bulk region by introducing a cutoff plane located at the inflection point of the average electrostatic potential difference $\tilde{\varphi}$, represented by the upper dashed curve in Fig.~\ref{fig3:alignment}(c). To describe the bending of the bands in the semiconductor, the electric field right below the surface is calculated and a Mott-Schottky extrapolation is performed to determine the position of Fermi energy deep inside the semiconductor according to the following equation:
\begin{equation}
    \epsilon_{F} = \tilde{\varphi_{0}} - e\Phi_{\rm FB}
\end{equation}
where the $\tilde{\varphi_{0}}$ stands for the asymptotic electrostatic potential that semi-infinitely extends in the bulk of the semiconductor and $\Phi_{\rm FB}$ is the flatband potential corresponding to the opposite Fermi energy of the neutral surface [Fig.~\ref{fig3:alignment}(a)]. Finally, the Fermi levels of the bulk semiconductor and of the interface are equilibrated by changing the distribution of charge between the explicit and implicit region while ensuring charge balance with the Helmholtz plane of counterions, leading to the equilibrated profile that is schematically depicted in Fig.~\ref{fig3:alignment}(d).

\begin{figure}
	\centering
	\includegraphics[width=0.45\textwidth]{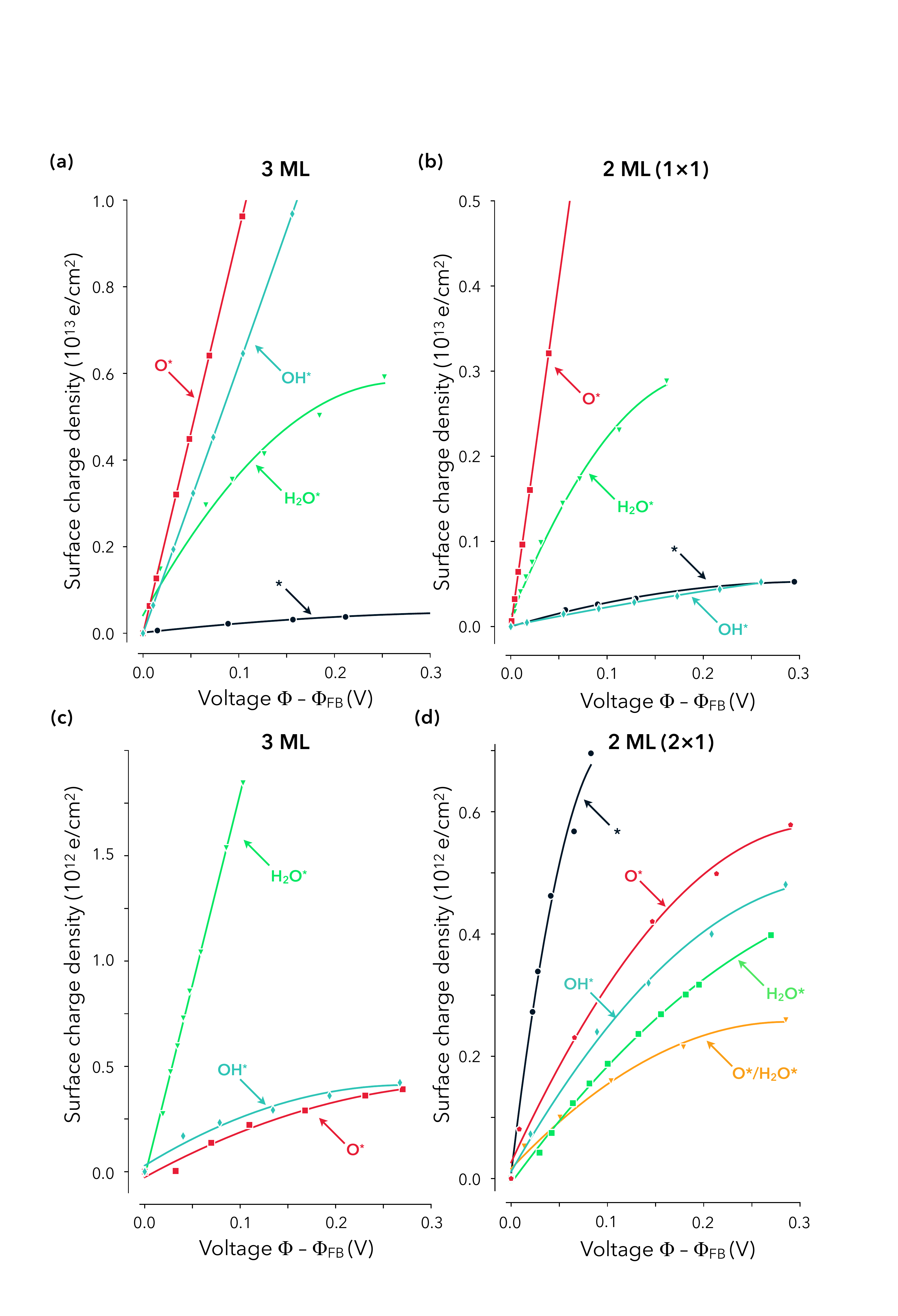}  
	\caption{\small Charge versus voltage characteristics of (a) 1 ML, (b) 2 ML (1$\times$1) (c) 3 ML and (d) 2 ML (2$\times$1) interfaces.}
	\label{fig4:CV-curves}
\end{figure}

The capacitive responses of the electrodes are obtained by varying the amount of charge added to the electrode (the opposite of the Helmholtz charge) \cite{PhysRevB.73.165402,PhysRevLett.110.086104}, and the resulting charge-voltage characteristics are reported in Fig.~\ref{fig4:CV-curves}.

These simulations show that the charge-voltage responses are critically dependent on the interface structures and terminations. More specifically, the bare 1 ML structure behaves similarly to an ideal semiconductor, with a limited amount of electronic charge trapped as surface states. This interface exhibits a moderate capacitive response across the low-voltage range, whereas an upshift of the capacitive response is observed for the hydrated interface 1 ML-H$_{2}$O$^*$, which indicates stronger charge trapping compared to the bare interface. Besides those two terminations, both the O$^*$ and OH$^*$ interfaces show substantial charge pinning, yielding a linear (ohmic) response. Such surface states reduce the extent of band bending, thus diminishing charge transport within the depletion region. It is interesting to note that, for the substantially reconstructed 3 ML surface, the O$^*$ and OH$^*$ terminations still maintain the semiconductor-like characteristics. This indicates that these surface structures could potentially provide a charge transport pathway for photocatalytic reactions.

Having determined the electrical characteristics of the proposed surface terminations, we now turn our attention to comparing their stability in electrolytic media in an effort to describe the voltage-induced reconstruction of the surface and its effects on photocatalytic durability and activity. Such surface-energy calculations are highly sensitive to the slab thickness and to the sampling of the Brillouin zone, potentially impacting the calculated surface energies. These sources of error can be eliminated by employing the methods proposed by Singh-Miller and Marzari \cite{Singh-Miller2009}. In this approach, the surface free energy $\gamma_{0}$ of a stoichiometric, symmetric slab is obtained as the limit
\begin{equation}
    \gamma_{0} = \lim_{N \to \infty} \frac{1}{2A_{s}} (E_{\rm slab}(N)-NE_{\rm bulk}),
    \label{subtract_surface_E}
\end{equation}
where $N$ is the number of slab layers, $E_{\rm slab}$ and $E_{\rm bulk}$ are the slab and bulk total energies, respectively, and $A_{s}$ stands for the surface area. In the limit of large $N$, Eq.~\ref{linear_surface_E} can be recast as
\begin{equation}
    E_{\rm slab}(N) = 2\gamma_0 A_{s} + NE_{\rm bulk},
    \label{linear_surface_E}
\end{equation}
reflecting the fact that the total energy of the slab should vary linearly as a function of the slab thickness with a slope equals to the bulk energy $E_{\rm bulk}$ of the material and an intercept corresponding to its surface energy $\gamma_0$.

\begin{figure}[t!]
	\centering
	\includegraphics[width=0.45\textwidth]{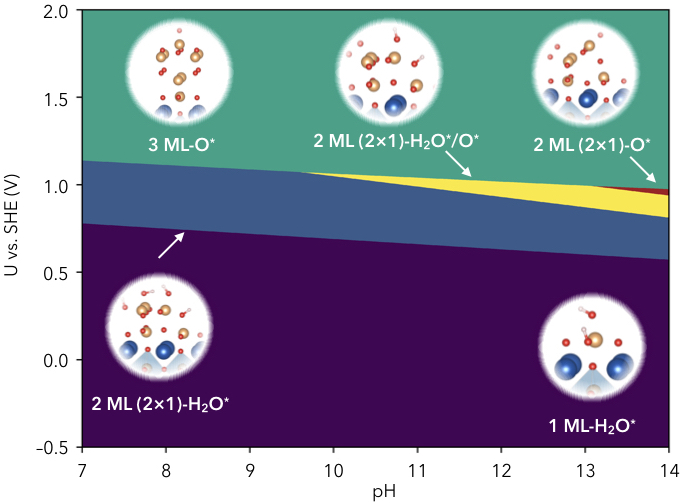}    
	\caption{\small Surface Pourbaix diagram showing a transition from the 1 ML to the 2 ML (2$\times$1) to the 3 ML terminations when applying an increasingly positive bias to the SrTiO$_{3}$ electrode.}
	\label{fig5:Surface_stability}
\end{figure}

We then consider the non-stoichiometry of the termination by taking into account the chemical potentials of the leaching and adsorbing elements in expressing the grand-canonical free energy $\gamma$ of the surface

\begin{equation}
    \gamma = \gamma_{0} - \mu_{\rm TiO_{2}} \Gamma_{\rm  TiO_{2}} - \mu_{\rm OH^{-}} \Gamma_{\rm OH^{-}} 
\end{equation}
in terms of the chemical potentials $\mu_{\rm TiO_{2}} = F(\rm TiO_{2})$ and $\mu_{\rm OH^{-}} = F({\rm {H_{2}O}}) - \frac{1}{2} F({\rm {H_{2\rm (g)}}}) + e{\rm \Phi_{\rm FB}} +k_{\rm B}T \rm{ln(10)} pH$, with the $\Gamma$'s and $F$'s being the surface densities and the calculated reference state energies of the surface species, respectively.  

Once the energy of the neutral interface is calculated, one obtains the free energy $\gamma^{*}$ of the charged interface from Lippmann's equation
\begin{equation}
    \gamma^{*} = \gamma - \int_{\Phi_{\rm FB}}^{\Phi}\sigma(\varphi)d\varphi,
\end{equation}
where $\Phi$ is the applied potential and $\sigma$ is the voltage-dependent surface charge density derived from the charge-voltage responses, leading in particular to significant changes in the surface stability of the more metallic (ohmic) interfaces at high voltage. The resulting stability analysis is reported in the surface Pourbaix diagram shown in Fig.~\ref{fig5:Surface_stability} for all the studied interfaces. At a pH of 14, corresponding to experimental alkaline conditions, calculations clearly indicate that the 3 ML-O$^*$ surface dominates at positive bias. This observation provides a direct first-principles confirmation that the 3 ML surface reconstruction is thermodynamically favorable under these operational conditions. Here, it is important to note that the solvation stabilizes the 2 ML (2$\times$1) interfaces compared to the 1 ML (2$\times$1)-H$_2$O$^*$ surface. In fact, under vacuum conditions, 1 ML (2$\times$1)-H$_2$O$^*$ exhibits higher stability, as shown in the vacuum Pourbaix diagram in Supporting Materials. Despite these variations, the predominant stability of 3 ML-O$^*$ interface is consistently confirmed by calculations in both solvation and vacuum environments. 

\begin{figure}[t]
	\centering
	\includegraphics[width=0.45\textwidth]{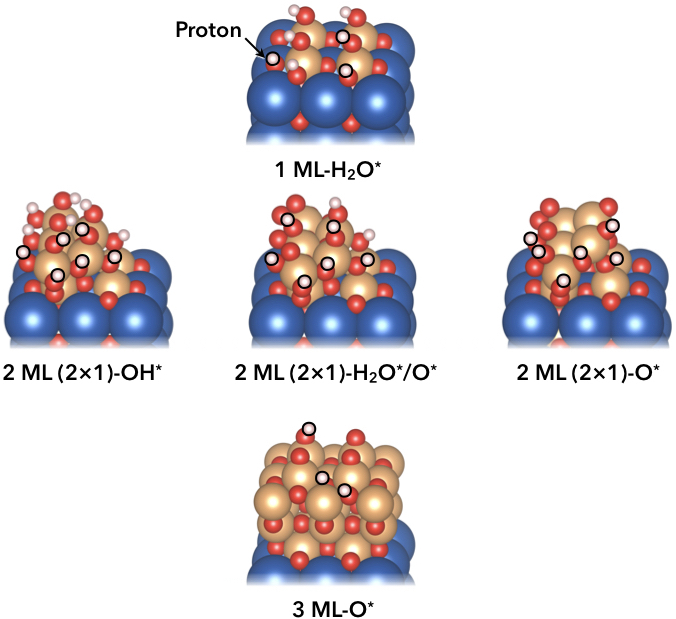} 
	\caption
	{\small Proton adsorption sites for the most stable surfaces in an alkaline solution. The stable reconstructed surfaces are 1 ML-H$_2$O$*$, 2 ML (2$\times$1) with its -OH$^*$, -H$2$O$*$, -O$^*$ interfaces and 3 ML-O$*$ with a focus on proton adsorption/desorption at all symmetrically unique hydroxyl groups and oxygens.}
	\label{fig:6}
\end{figure}

To conclude the discussion, we study the influence of the reconstructed surface structure of SrTiO$_{3}$ on the theoretical hydrogen evolution overpotential. We consider the most stable structures (1 ML-H$_2$O$^*$, 2 ML (2$\times$1)-OH$^*$, H$_2$O$^*$/O$^*$, O$^*$ and 3 ML-O$^*$) based on the results reported in Fig.~\ref{fig5:Surface_stability} and systematically examine the proton adsorption sites shown in Fig.~\ref{fig:6}.
 
\begin{figure}[t]
	\centering
	\includegraphics[width=0.47\textwidth]{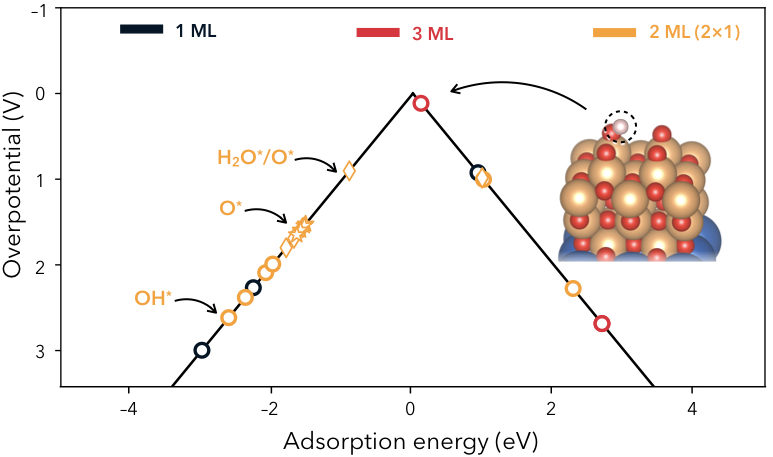} 
	\caption
	{\small Thermodynamic volcano plot for hydrogen evolution on 1 ML-H$_2$O, 2 ML (2$\times$1)-OH$^*$, H$_2$O$^*$/O$^*$, O$^*$ interfaces and 3 ML-O$^*$ reconstructed SrTiO$_3$ surfaces. For the 2 ML (2$\times$1) terminations, the OH$^*$, H$_2$O$^*$/O$^*$, O$^*$ terminations are labeled with circle, diamond and star markers, respectively.}
	\label{fig7: Volcano}
\end{figure}

The adsorption energy $\Delta F_{\rm H}$ is expressed as
\begin{equation}
    \Delta F_{\rm H} = F_{\rm H^*} - F_*-\frac{1}{2} F_{\rm H_{2}},
\end{equation}
 where $F_{\rm H^*}$ is the energy of the proton adsorbed on the SrTiO$_{3}$ surface and $F_*$ is the total energy of the adsorption site. The adsorption free energy of $\rm H^*$ is then calculated from
\begin{equation}
    \Delta G_{\rm H} = \Delta F_{\rm H} + \Delta{\rm ZPE}_{\rm H} - T\Delta S_{\rm H},
\end{equation}

\begin{equation}
    \Delta {\rm ZPE}_{\rm H} = {\rm ZPE}_{\rm H^*} - \frac{1}{2} {\rm ZPE_{H_{2}(g)}},
\end{equation}
where $\Delta$ZPE$_{\rm H}$ is the change of zero-point vibrational energy of hydrogen, and $\Delta S_{\rm H}$ is the change in entropy upon the adsorption. We compute ZPE$_{\rm H^*}$ at the interfaces using $\Gamma$-point phonon calculations. The zero-point energy of reference hydrogen is computed in the gas phase as ZPE$_{\rm H_2(g)}$. The entropy contributions are obtained from experimental data at 300 K \cite{doi:10.1021/ci010445m}. Based on these calculations, we construct the volcano plot shown in Fig.~\ref{fig7: Volcano}, where the peak stands for the thermodynamic conditions most favorable to hydrogen evolution, according to the Sabatier principle \cite{doi:10.1021/ja0504690,C5EE02179K}.
\vspace{1cm}

The volcano plot exhibits large changes in the overpotentials for different surface structures and clearly indicates that the outermost oxygen site from the 3 ML-O$^*$ interface is the closest to the volcano peak, with an overpotential of 0.11 V. All other terminations tend to adsorb the protons either too strongly or too weakly. Comparing these results to the previous charge-voltage curves and surface reconstruction diagram (Fig.~\ref{fig4:CV-curves} and Fig.~\ref{fig5:Surface_stability}), the 3 ML-O$^*$ surface is anticipated to present semiconductor-like interfacial charge transport characteristics, to dominate the surface stability and to be active for hydrogen evolution, providing a quantitative interpretation of its unexpectedly high photocatalytic activity upon anodic preparation.

%%%%%%%%%%%%%%%%%%%%%%%%%%%%%%%%%%%%%%%%%%%%
\section{Conclusion}                                         
\label{sec:conclusion}                                       
%%%%%%%%%%%%%%%%%%%%%%%%%%%%%%%%%%%%%%%%%%%%

In this work, we have investigated the surface structures of SrTiO$_{3}$ under electrochemical conditions. By applying an embedded quantum-mechanical model based on the self-consistent continuum solvation approach, we have determined the electrical response of the reconstructed interfaces of SrTiO$_{3}$ from first principles. We have shown that the surface terminations strongly affect the electrification of SrTiO$_{3}$ photoelectrodes, leading to a variety of interfacial charge-voltage behaviors ranging from ohmic to semiconducting. The 3 ML termination has been found to be among the surface structures that provide potential charge transport pathways. We have then computed the surface free energies of all interfaces under applied potential in alkaline solutions. These electrochemical calculations have suggested that the 3 ML-O$^*$ is the most stable of the structures considered under positive bias. The catalytic activity of the 3 ML termination has been estimated to be the strongest for the hydrogen evolution reaction based on vacuum calculations of hydrogen binding energies. Our calculations provide direct molecular insights into the voltage-induced formation of the triple-TiO$_2$ termination and into the beneficial influence of this reconstruction on the photocatalytic activity of SrTiO$_{3}$.

%%%%%%%%%%%%%%%%%%%%%%%%%%%%%%%%%%%%%%%%%%%%
\section*{Acknowledgments}
\label{sec:acknoledgments}
%%%%%%%%%%%%%%%%%%%%%%%%%%%%%%%%%%%%%%%%%%%%
The authors acknowledge financial support from the National Science Foundation under grant number DMREF-1729338. The computational work was performed using high-performance computing resources from the Penn State Institute of CyberScience. The authors would like to thank Iurii Timrov, Matteo Cococcioni and Nicola Marzari for making available the preliminary density-function perturbation implementation of the linear-response Hubbard $U$ correction and for fruitful discussions and suggestions.

\bibliographystyle{apsrev4-1}

%merlin.mbs apsrev4-1.bst 2010-07-25 4.21a (PWD, AO, DPC) hacked
%Control: key (0)
%Control: author (72) initials jnrlst
%Control: editor formatted (1) identically to author
%Control: production of article title (-1) disabled
%Control: page (0) single
%Control: year (1) truncated
%Control: production of eprint (0) enabled
%

%\bibliography{STO_Band_Bending}

\balance

\end{document}